\documentclass[12pt]{article}
\catcode`\@=11
\@addtoreset{equation}{section}

\global\arraycolsep=2pt
\usepackage{mathrsfs,amsbsy,amssymb,latexsym,amsfonts,amsmath,cite}
\usepackage{graphicx,color}

\newcommand\no{\nonumber}

\newcommand{\dd}{{\rm d}}

\allowdisplaybreaks

\begin{document}

\begin{flushright}
\parbox{4cm}
{KUNS-2672 \\ 
}
\end{flushright}

\vspace*{0.5cm}

\begin{center}
{\Large \bf 
Comments on 2D dilaton gravity system \\ with a hyperbolic dilaton potential}
\vspace*{1.5cm}\\
{\large Hideki Kyono\footnote{E-mail:~h\_kyono@gauge.scphys.kyoto-u.ac.jp}, 
Suguru Okumura\footnote{E-mail:~s.okumura@gauge.scphys.kyoto-u.ac.jp}
and Kentaroh Yoshida\footnote{E-mail:~kyoshida@gauge.scphys.kyoto-u.ac.jp}} 
\end{center}

\vspace*{0.5cm}

\begin{center}
{\it Department of Physics, Kyoto University, \\ 
Kitashirakawa Oiwake-cho, Kyoto 606-8502, Japan} 
\end{center}

\vspace{1cm}

\begin{abstract}
We proceed to study a (1+1)-dimensional dilaton gravity system with a hyperbolic 
dilaton potential. Introducing a couple of new variables leads to  
two copies of Liouville equations with two constraint conditions.  
In particular, in conformal gauge, the constraints can be expressed with Schwarzian derivatives. 
We revisit the vacuum solutions in light of the new variables  
and reveal its dipole-like structure. 
Then we present a time-dependent solution 
which describes formation of a black hole with a pulse. Finally, the black hole thermodynamics 
is considered by taking account of conformal matters from two points of view: 
1) the Bekenstein-Hawking entropy and 2) the boundary stress tensor. 
The former result agrees with the latter one with a certain counter-term.  
\end{abstract}

\setcounter{footnote}{0}
\setcounter{page}{0}
\thispagestyle{empty}

\newpage

\tableofcontents

\section{Introduction}

The AdS/CFT correspondence \cite{M,GKP,Witten} has been recognized as a realization 
of the holographic principle \cite{Hooft,Susskind}. However, the rigorous proof has not 
been provided yet, although the integrable structure behind the correspondence 
has led to great advances along this direction (For a comprehensive review see \cite{Beisert}). 
A recent interest in the study of AdS/CFT is to construct a toy model which realizes  
a holographic principle at the full quantum level. 
Recently, Kitaev proposed an intriguing model \cite{Kitaev} 
as a variant of the Sachdev-Ye (SY) model \cite{SY}. 
This is a one-dimensional quantum-mechanical system 
composed of $N \gg 1$ Majorana fermions with a random, all-to-all quartic interaction. 
This model is now referred to as the Sachdev-Ye-Kitaev (SYK) model. 
For some recent progress, see 
\cite{PR,MS,Jensen,GR,FGMS,Witten2,many,Klebanov,Nishinaka,Garcia,SW,Itoyama}.

\medskip 

A possible candidate of the gravity dual for the SYK model is a 1+1 dimensional 
dilaton gravity system with a certain dilaton potential (For a nice review see \cite{V}). 
This model was originally introduced by Jackiw \cite{Jackiw} and Teitelboim \cite{Teitelboim}.  
Then it has been further studied by Almheiri and Polchinski \cite{AP} in light of holography 
\cite{EMV,MSY,Cvetic}. This model contains interesting solutions like renormalization group flow 
solutions, black holes, time-dependent solutions which describe formation of a black hole. 
Since the black hole is asymptotically AdS$_2$\,, the boundary stress tensor 
computed in the standard manner leads to the associated entropy, 
which agrees with the Bekenstein-Hawking entropy. 

\medskip 

In the preceding work \cite{KOY}, we have studied deformations of this dilaton gravity system 
by employing a Yang-Baxter deformation technique \cite{Klimcik,DMV,MY-YBE}. 
The dilaton potential is deformed from a simple quadratic form to 
a hyperbolic function-type potential.
We have presented the vacuum solutions and studied the associated geometries. 
As a remarkable feature, the UV region of the geometries is universally deformed to 
dS$_2$ and a new naked singularity is developed\footnote{It is worth noting that 
an interesting work along a similar line has been done in \cite{Hofman}, 
in which the UV region is AdS$_2$ 
and the IR region is deformed to dS$_2$\,. }. 
The vacuum solutions include a deformed black hole solution, 
which reduces to the original solution \cite{AP} in the undeformed limit.
We have computed the entropy of the deformed black hole by evaluating the boundary stress tensor 
with a certain counter-term. 
The resulting entropy still agrees with the Bekenstein-Hawking entropy.   

\medskip 

In this paper, we will further study the dilaton gravity system with the hyperbolic 
dilaton potential. Introducing a couple of new variables leads to two copies of Liouville equations 
with two constraint conditions. As a remarkable feature, the constraints can be expressed 
in terms of Schwarzian derivatives. The new variables are so powerful to study solutions 
and enable us to reveal the dipole-like structure of the vacuum solutions. 
As a benefit, we present a time-dependent solution 
which describes formation of a black hole with a pulse. Finally, the black hole thermodynamics 
is considered by taking account of conformal matters from two points of view: 
1) the Bekenstein-Hawking entropy and 2) the boundary stress tensor. 
The former result agrees with the latter one with a counter-term modified in a certain way.  

\medskip 

This paper is organized as follows. In section 2, we give a short review of 
the deformed dilaton gravity system. Then we revisit the vacuum solutions 
by introducing a couple of new variables. 
In section 3, we consider how to treat matter fields and  derive 
a time-dependent solution which describes formation of a black hole with a pulse. 
In section 4, adding conformal matters, we derive a deformed black hole solution.  
Then we reproduce the Bekenstein-Hawking entropy by computing the boundary 
stress tensor with a certain counter-term.  
Section 5 is devoted to conclusion and discussion.

\section{Hyperbolic dilaton potential and new variables}
\label{section2}

In this section, we revisit a (1+1)-dimensional dilaton gravity system 
with a hyperbolic dilaton potential. This system has been introduced in \cite{KOY} 
as a Yang-Baxter deformation \cite{Klimcik,DMV,MY-YBE} 
of the Jackiw-Teitelboim (JT) model \cite{Jackiw,Teitelboim}. 

\subsection{A dilaton gravity system with a hyperbolic potential}

In the following, we will work in the Lorentzian signature and 
the (1+1)-dimensional spacetime is described by the coordinates 
$x^{\mu}=(t,x)~(\mu=0,1)$\,. This system contains the metric $g_{\mu\nu}$ and 
the dilaton $\Phi$ as the basic ingredients. We may add 
other matter fields but will not do that in section \ref{section2}.

\medskip 

The classical action for $g_{\mu\nu}$ and $\Phi$ 
is given by \cite{KOY}
\begin{eqnarray}
S_{\Phi } &=& \frac{1}{16\pi G}\int\!\dd^2 x \sqrt{-g}\left[\Phi^2 R + \frac{1}{\eta}
\sinh\left( 2 \eta \Phi^2 \right)\right] 
+\frac{1}{8 \pi G}\int\!\dd t \sqrt{-\gamma_{tt}}\,\Phi^2 K\,, 
\label{action}
\end{eqnarray}
where $G$ is a two-dimensional Newton constant, and $R$ and $g$ are Ricci scalar 
and determinant of $g_{\mu\nu}$\,. 
The last term is the Gibbons-Hawking term that contains an extrinsic metric $\gamma_{tt}$ 
and an extrinsic curvature $K$\,. A remarkable point of this action is the second term. 
This is a dilaton potential of hyperbolic function, where $\eta$ is a real 
constant parameter\footnote{
In this paper, we slightly changed the normalization of the dilaton potential
from \cite{KOY}.
Therefore the constant factors of solutions are also changed.}. 

\medskip

In the $\eta \to 0$ limit, the classical action (\ref{action}) reduces to 
the JT model (without matter fields)
\begin{eqnarray}
S_{\Phi } &=& \frac{1}{16\pi G}\int\!\dd^2 x \sqrt{-g}\Bigl[\Phi^2 R +  2 \Phi^2 \Bigr] 
+\frac{1}{8 \pi G}\int\!\dd t \sqrt{-\gamma_{tt}}\,\Phi^2 K\,.
\end{eqnarray}
Thus the classical action (\ref{action}) can be regarded as a deformation of the JT model.

\subsubsection*{The vacuum solutions}

The deformed model (\ref{action}) gives rise a three parameter family of vacuum solutions\footnote{Here the dilaton 
is turned on, but the solution is still called ``vacuum'' solution, according to the custom. }  
\cite{KOY}, 
\begin{eqnarray}
\dd s^2 &=&\frac{1+\eta^2P^2}{1-\eta^2 (X\cdot P)^2}\frac{-\dd t^2+ \dd z^2}{z^2} \,, \label{2.3}\\ 
\Phi^2 &=&\frac{1}{2\eta}\log\left|\frac{1+\eta (X\cdot P)}{1-\eta (X\cdot P)}\right|\,, 
\label{2.4}
\end{eqnarray}
where $X$ and $P$ are defined as 
\begin{eqnarray}
X^I &\equiv& \frac{1}{z}\left(t~,~\frac{1}{2}(1+t^2-z^2)~,~\frac{1}{2}(1-t^2+z^2)\right)\,,\no\\
P_I &\equiv&(\beta~,~\alpha-\gamma~,~\alpha+\gamma)\, \quad \qquad (I,J=1,2,3)\,,
\end{eqnarray}
and the products $X\cdot P$ and $P^2$ are given by 
\begin{eqnarray}
X \cdot P &\equiv& X^I P_I = \frac{\alpha+\beta \, t+\gamma\,(-t^2+z^2)}{z}\,,\no\\
P^2&=& \eta^{IJ}P_IP_J=-(\beta^2+4\alpha\gamma)\,. 
\end{eqnarray}
Here the metric of the embedding space $\mathbb{M}^{2,1}$ is taken 
as $\eta_{IJ}=\text{diag}(-1,1,-1)$\,. 
This family labeled by $\alpha$, $\beta$ and $\gamma$ is associated with the most general Yang-Baxter deformation. 
In other words, the effect of Yang-Baxter deformation appears only through the factor 
$\eta (X\cdot P)$\,.  
It should also be remarked that a black hole solution is contained as a special case \cite{KOY}. 

\medskip 

It is worth noting that the conformal factor of the metric (\ref{2.3}) can 
be expressed as a Schwarzian derivative of the dilaton (\ref{2.4}) like 
\begin{eqnarray}
\sqrt{\,\frac{1}{2}{\rm Sch}\{\Phi^2, \,\eta (X\cdot P) \}}\, 
= \frac{z^2}{1+\eta^2 P^2}\,{\rm e}^{2\omega}\, \,.  
\end{eqnarray}
Here ${\rm Sch}\{ X, x\}$ denotes the Schwarzian derivative defined as 
\begin{eqnarray}
{\rm Sch}\{ X, x \} \equiv \frac{ X\/'''}{X\/'} - \frac{3}{2} \left( \frac{ X\/'' }{ X\/' } \right)^2\,. 
\end{eqnarray}

\subsection{Introducing a couple of new variables}

Let us first rewrite the metric into the following form: 
\begin{eqnarray}
\dd s^2 &=& {\rm e}^{2 \omega}\tilde{g}_{\mu\nu}\dd x^\mu \dd x^\nu \,. 
\end{eqnarray}
Then the classical action (\ref{action}) can be rewritten as 
\begin{eqnarray}
S_{\Phi } &=& \frac{1}{16\pi G}\int\!\dd^2 x \sqrt{-\tilde{g}}\left[\Phi^2 \tilde{R} 
+2\tilde{\nabla}\Phi \tilde{\nabla}\omega+ \frac{{\rm e}^{2\omega}}{\eta}
\sinh\left( 2 \eta \Phi^2 \right)\right]\,. \label{action1}
\end{eqnarray}
In order to simplify this expression, it is helpful to introduce a couple of new valuables:
\begin{eqnarray}
\omega_{1} \equiv \omega +   \eta \Phi^2\,, \qquad 
\omega_{2} \equiv \omega -  \eta \Phi^2 \,.  
\label{defomega}
\end{eqnarray}
Then the action (\ref{action1}) becomes the sum of 
two Liouville systems: 
\begin{eqnarray}
S_{\Phi } &=& \frac{1}{32\pi G \eta}\int\!\dd^2 x \sqrt{-\tilde{g}}\left[ \left(\omega_1 \tilde{R} 
+(\tilde{\nabla}\omega_1)^2+ {\rm e}^{2\omega_1}\right) 
- \left(\omega_2 \tilde{R} +(\tilde{\nabla}\omega_2)^2 
+ {\rm e}^{2\omega_2}\right) \right] \, \label{2.12}\\
&\equiv&S_{\omega_1}+S_{\omega_2}\,. \nonumber 
\end{eqnarray}
By taking variations of the action (\ref{2.12}) with respect to $\omega_1$ and $\omega_2$\,, 
it is easy to derive the following equations of motion:
\begin{eqnarray}
\tilde{R}-2(\tilde{\nabla}\omega_1)^2+2{\rm e}^{2\omega_1}&=&0\,, \no\\
\tilde{R}-2(\tilde{\nabla}\omega_2)^2+2{\rm e}^{2\omega_2}&=&0\,. 
\end{eqnarray}
Taking a variation with $\tilde{g}_{\mu\nu}$ gives rise to the constraints 
\begin{eqnarray}
\tilde{T}^{(1)}_{\mu\nu} +\tilde{T}^{(2)}_{\mu\nu}=0 \,, 
\end{eqnarray}
where $\tilde{T}^{(1)}_{\mu\nu}$ and  $\tilde{T}^{(2)}_{\mu\nu}$ are the energy-momentum 
tensors defined as, respectively, 
\begin{eqnarray}
\tilde{T}^{(1)}_{\mu\nu} \equiv \frac{-2}{ \sqrt{-\tilde{g}} } 
\frac{\delta S_{\omega_1} }{ \delta \tilde{g}^{\mu\nu}} \,,\qquad
\tilde{T}^{(2)}_{\mu\nu} \equiv \frac{-2}{ \sqrt{-\tilde{g}} } 
\frac{\delta S_{\omega_2} }{ \delta \tilde{g}^{\mu\nu}} \,, 
\end{eqnarray}
and the explicit forms are given by 
\begin{eqnarray}\label{EM-tensor}
\tilde{T}^{(1)}_{\mu\nu}
&=&\frac{1}{16 \pi G \eta} 
\left[ \tilde{\nabla}_\mu \tilde{\nabla}_\nu\omega_1 -(\tilde{\nabla}_\mu \omega_1)
( \tilde{\nabla}_\nu \omega_1 )-\frac{1}{2}\tilde{g}_{\mu\nu}(2 \tilde{\nabla}^2\omega_1 
-  (\tilde{\nabla}\omega_1)^2 - {\rm e}^{2 \omega_1}) \right]\,, \no\\
\tilde{T}^{(2)}_{\mu\nu}
&=&\frac{-1}{16 \pi G \eta} 
\left[ \tilde{\nabla}_\mu \tilde{\nabla}_\nu\omega_2 -(\tilde{\nabla}_\mu \omega_2)
( \tilde{\nabla}_\nu \omega_2 )-\frac{1}{2}\tilde{g}_{\mu\nu}(2 \tilde{\nabla}^2\omega_2 
-  (\tilde{\nabla}\omega_2)^2 - {\rm e}^{2 \omega_2}) \right]\,. 
\end{eqnarray}
Thus, by employing the new variables $\omega_1$ and $\omega_2$\,, 
the deformed system (\ref{action}) has been simplified drastically.

\subsection{Conformal gauge and Schwarzian derivatives}

In the following, we will work with the usual conformal gauge 
\begin{eqnarray}
\dd s^2 &=& - {\rm e}^{2 \omega}\dd x^+ \dd x^- \qquad (x^\pm \equiv t\pm z)\,. 
\end{eqnarray}
Then the equations of motion obtained from (\ref{action}) are given by 
\begin{eqnarray}
4\partial_+\partial_-\Phi^2 +\frac{1 }{\eta} {\rm  e}^{2\omega} \sinh\left( 2 \eta \Phi^2\right) 
&=& 0\,,\label{originaleom1} \\
4\partial_+\partial_-\omega + {\rm e}^{2\omega} \cosh\left( 2 \eta \Phi^2\right) 
&=&0 \,, \label{originaleom2}\\
- {\rm e}^{2\omega}\partial_+ ({\rm e}^{-2\omega}\partial_+\Phi^2) 
&=&0\,, \label{const+} \,\\ 
- {\rm e}^{2\omega}\partial_- ({\rm e}^{-2\omega}\partial_-\Phi^2)&=&0\,.\label{const-}
\end{eqnarray} 
By solving the above equations, the general vacuum solution has been discussed 
in \cite{KOY}. However, as we will show below, the deformed model (\ref{action})
 has a nice property, 
 with which we can discuss classical solutions in a more systematic way.

\subsubsection*{New variables revisited}

In conformal gauge, the classical action for $\omega_1$ and $\omega_2$ is further simplified as 
\begin{eqnarray}
S_{\Phi } &=& \frac{1}{8 \pi G \eta}\int\!\dd^2 x \left[ \left( -\partial_+\omega_1\, 
\partial_-\omega_1 + \frac{1}{4} {\rm e}^{2\omega_1} \right) 
- \left(-\partial_+\omega_2\, \partial_-\omega_2 
+ \frac{1}{4} {\rm e}^{2\omega_2} \right) \right] \,. 
\end{eqnarray}
The equations of motion take the standard forms of the Liouville equation
\begin{eqnarray}
4\partial_+\partial_- \omega_{1} + {\rm e}^{2\omega_{1} } =0 \,, \qquad 
4\partial_+\partial_- \omega_{2} + {\rm e}^{2\omega_{2} } = 0 \,. \label{eom1}
\end{eqnarray}
The general solutions of Liouville equation are given by\footnote{A similar structure 
was found in a two-dimensional non-linear sigma model whose target space 
is given by a three-dimensional Schr$\ddot{\rm o}$dinger spacetime \cite{Sch-dipole}. } 
\begin{eqnarray}
{\rm e}^{2 \omega_{1} } 
= \frac{4 \partial_+ X^+_1 \partial_- X^-_1 }{  \left( X^+_1 - X^-_1 \right)^2}\,, 
\qquad 
{\rm e}^{2 \omega_{2} } 
= \frac{4 \partial_+ X^+_2 \partial_- X^-_2 }{  \left( X^+_2 - X^-_2 \right)^2}\,\,, 
\label{Liouville}
\end{eqnarray}
where $X^+_i=X^+_i(x^+)$  and $X^-_i=X^-_i(x^-)$ are arbitrary holomorphic 
and anti-holomorphic functions, respectively. 

\medskip

Note that the equations (\ref{eom1}) can be expressed by using the metric and dilaton.
\begin{eqnarray}
4\partial_+\partial_- (\omega+\eta\Phi) + {\rm e}^{2\omega+2\eta \Phi } =0 \,,\no\\
4\partial_+\partial_- (\omega-\eta\Phi) + {\rm e}^{2\omega-2\eta\Phi } = 0 \,.
\end{eqnarray}
By summing and subtracting them each other, 
the equations of motion (\ref{originaleom1}) and (\ref{originaleom2}) can be reproduced.

\medskip

By taking $\tilde{g}_{\mu\nu}=\eta_{\mu\nu}$ in \eqref{EM-tensor},
the energy-momentum tensors are also rewritten as 
\begin{eqnarray}
\tilde{T}^{(1)}_{+-} +\tilde{T}^{(2)}_{+-}&=&\frac{-1}{16\pi G \eta }(\partial_+\partial_-\omega_1 
+\frac{1}{4}{\rm e}^{2\omega_1}-\partial_+\partial_-\omega_2 
- \frac{1}{4}{\rm e}^{2\omega_2})\,, \label{c1}\\
\tilde{T}^{(1)}_{\pm\pm} +\tilde{T}^{(2)}_{\pm\pm} &=& 
\frac{1}{16\pi G \eta }( \partial_\pm \partial_\pm \omega_1 
- \partial_\pm \omega_1 \partial_\pm \omega_1 
- \partial_\pm \partial_\pm \omega_2 
+ \partial_\pm \omega_1 \partial_\pm \omega_1 )\,.  \label{c2}
\end{eqnarray}
The first condition (\ref{c1}) vanishes automatically due to the equations of motion. 
Two conditions in (\ref{c2}) give rise to nontrivial constraints for the solutions 
of the equations of motion (\ref{eom1}). By using the definitions of $\omega_1$ and $\omega_2$ 
in (\ref{defomega})\,, it is easy to directly see that the constraints 
in (\ref{c1}) and (\ref{c2}) are equivalent to the ones in (\ref{const+}) and (\ref{const-})\,.

\medskip 

By using the general solutions (\ref{Liouville}), the constraint conditions 
for the  holomorphic (antiholomorphic) functions $X^+_i$ ($X^-_i$) can be rewritten as 
\begin{eqnarray}\label{constraint-sch}
{\rm Sch}\{ X^+_1, x^+\} - {\rm Sch}\{ X^+_2, x^+\}  &=&0\,, \no\\
{\rm Sch}\{ X^-_1, x^-\} - {\rm Sch}\{ X^-_2, x^-\}  &=&0\,. 
\end{eqnarray}
These constraints mean that the  holomorphic (antiholomorphic) functions 
should be the same functions, up to linear fractional transformations 
\begin{eqnarray}\label{LFT}
X^+_2(x^+) &=&  \frac{ a X^+_1 + b }{ c X^+_1 +d } \qquad (a,b,c,d \in \mathbb{R})\,, \no \\
X^-_2(x^-) &=& \frac{ a\/' X^-_1 + b\/' }{ c\/' X^-_1 +d\/' } \qquad 
(a',b',c',d' \in \mathbb{R})\,. 
\end{eqnarray}
Because ${\rm e}^{2 \omega_{1} }>0$ and ${\rm e}^{2 \omega_{2} }>0$\,,  
determinants of the transformations must be positive: 
\begin{eqnarray}\label{constraint-abcd}
ad-bc>0\,,\qquad a\/'d\/'-b\/'c\/'>0\,. 
\end{eqnarray}
This ambiguity comes from the appearance of Schwarzian derivatives.

\subsection{Vacuum solutions revisited}

In this subsection, let us revisit the vacuum solutions by employing a couple of the new 
variables (\ref{defomega})\,. 
Before going to the detail, it is helpful to recall that 
the original metric and dilaton can be reconstructed from $\omega_1$ and $\omega_2$ 
through the following relations:  
\begin{eqnarray}\label{relation12}
{\rm e}^{ 2 \omega } = \sqrt{{\rm e}^{ 2\omega_{1}} {\rm e}^{2 \omega_{2 } }}\,, \qquad 
\Phi^2 = \frac{1}{2 \eta} ( \omega_{1} -\omega_{2 } )=\frac{1}{4 \eta}
\log \left(\frac{{\rm e}^{2\omega_1}}{{\rm e}^{2\omega_2}}\right)\,. 
\end{eqnarray} 

\medskip 

Here let us take a parametrization for the linear fractional transformations,  
which come from \eqref{constraint-sch} as follows:\footnote{
Note that we can take this parametrization without loss of generality.}
\begin{eqnarray}
X^+_1(x^+) &=& \frac{ (1-\eta \beta)X^+(x^+) - 2 \eta \alpha }{-2 \eta \gamma X^+(x^+) 
+ (1+\eta \beta) } \,, 
\qquad X^-_1(x^-)  = X^-(x^-)\,, \no \\
X^+_2(x^+) &=& \frac{ (1+\eta \beta)X^+(x^+) + 2\eta \alpha }{2\eta \gamma X^+(x^+) 
+ (1-\eta \beta) }\,,  
\qquad X^-_2(x^-) = X^-(x^-)\,. 
\label{YB form}
\end{eqnarray}
Because of the constraint \eqref{constraint-abcd}, we have to work 
in a restricted parameter region with 
\begin{eqnarray}\label{constraint-albega}
1-\eta^2(\beta^2 +4 \alpha \gamma) >0\,. 
\end{eqnarray}
Then the solutions in (\ref{Liouville}) are expressed as  
\begin{eqnarray}
{\rm e}^{2 \omega_{1} }=\frac{4 \left( 1-\eta^2(\beta^2 +4 \alpha \gamma) \right)
\partial_+ X^+\partial_- X^- }{ 
\left( X^+ -X^- - \eta( 2 \alpha +\beta (X^+ +X^-) -2\gamma X^+ X^-) \right)^2 }\,, \no\\
{\rm e}^{ 2 \omega_{2} }=\frac{4\left( 1-\eta^2(\beta^2 +4 \alpha \gamma) \right) 
\partial_+ X^+\partial_- X^-}{ 
\left( X^+ -X^- + \eta(2 \alpha +\beta (X^+ +X^-) -2\gamma X^+ X^-) \right)^2 }\,. 
\end{eqnarray}
Thus the general solution of $\omega$ and $\Phi^2$ are also determined  
through the relation \eqref{relation12}. 
Given that $X^\pm(x^\pm)=x^\pm$\,, the deformed metric and 
dilaton become\footnote{
Here the condition \eqref{constraint-albega} is consistent with 
the positivity of ${\rm e}^{2 \omega_1}$ and ${\rm e}^{2 \omega_2}$\,.}
\begin{eqnarray}
{\rm e}^{2\omega}&=&\frac{ 1-\eta^2(\beta^2 +4 \alpha \gamma) }{ 
z^2 - \eta^2 \left( \alpha +\beta t +\gamma (-t^2+z^2) \right)^2  }\,, \no\\\no\\
\Phi^2&=&
\frac{1}{2\eta}{\text{log}}\left|\frac{ z + \eta( \alpha +\beta t +\gamma (-t^2+z^2)   }{  
z - \eta( \alpha +\beta t +\gamma (-t^2+z^2) }\right|\,. 
\end{eqnarray}
This metric is the same as the result obtained in \cite{KOY} 
as a Yang-Baxter deformation of AdS$_{2}$\,, up to a scaling factor.

\medskip 

For concreteness, let consider a simple case of (\ref{YB form}) 
with $\alpha =1,~\beta=\gamma=0$\,.
Then conformal factors of the metrics 
for $X_1$ and $X_2$ are given by, respectively,  
\begin{eqnarray}
{\rm e}^{2\omega_1} = \frac{1}{(z-\eta)^2}\,,  \qquad 
{\rm e}^{2\omega_2} = \frac{1}{(z+\eta)^2}\,. 
\end{eqnarray}
For each of the AdS$_2$ factors, the origin of the $z$-direction is shifted by $\pm \eta$\,. 
Another example is the case with $\alpha=1/2,~\beta=0,~\gamma=\mu/2$ 
(where $\mu$ is a positive), in which 
we have considered a deformed black hole solution \cite{KOY}\footnote{
Note that for arbitrary values of $\alpha$, $\beta$ and $\gamma$\,, 
black hole solutions can be realized by employing the following coordinate transformation, 
\begin{eqnarray}
X^{\pm}(x^\pm)=\frac{\sqrt{\mu}}{2 \gamma}\tanh \left(\sqrt{\mu}\, 
x^\pm\right)+\frac{\beta}{2 \gamma}\qquad (\mu=\beta^2+4\alpha\, \gamma)\,.
\end{eqnarray}}.

\section{Solutions with matter fields}

In this section, we shall include additional matter fields. 
Then the action is given by a sum of the dilaton part $S_{\Phi}$ 
and the matter part $S_{\rm matter}$ like 
\begin{eqnarray}
S &=&S_{\Phi}+S_{\rm matter}\,.
\end{eqnarray}
Note here that we have not specified the concrete expression of 
the matter action $S_{\rm matter}$ yet.  In general, $S_{\rm matter}$ may depend on the metric, 
dilaton as well as additional matter fields. 
Hence the inclusion of matter fields leads to the modified equations: 
\begin{eqnarray}
4\partial_+\partial_-\Phi^2 +\frac{1 }{\eta}  {\rm e}^{2\omega} \sinh\left( 2 \eta \Phi^2\right)
&=& 32 \pi G\, T_{+-}\,,  \no\\
4\partial_+\partial_-\omega + {\rm  e}^{2\omega} 
\cosh\left( 2 \eta \Phi^2\right) &=& 
- 16 \pi G\, \frac{\delta S_{\rm matter}}{\delta \Phi^2} \,,  \no\\
- {\rm e}^{2\omega}\partial_+({\rm e}^{-2\omega}\partial_+\Phi^2) 
&=& 8 \pi G\, T_{++}\,, \no\\ 
- {\rm e}^{2\omega}\partial_-({\rm e}^{-2\omega}\partial_-\Phi^2) 
&=& 8 \pi G\, T_{--}\,. 
\label{matter eq}
\end{eqnarray}
Furthermore, one needs to take account of the equation of motion 
for the matter fields, which is provided as the conservation law of the energy-momentum 
tensor $T_{\mu\nu}$ defined as
\begin{eqnarray}
T_{\mu\nu}&\equiv&-\frac{2}{\sqrt{-g}}\frac{\delta S_{\rm matter}}{\delta g^{\mu\nu}}\,. 
\end{eqnarray}
So far, it seems difficult to treat the general expression of $T_{\mu\nu}$\,. 
Hence we will impose some conditions for $T_{\mu\nu}$ hereafter.

\subsection{A certain class of matter fields}

For simplicity, 
let us consider a certain class of matter fields 
by supposing the following properties:  
\begin{eqnarray}
T_{+-}=0\,,  \qquad \frac{\delta S_{\rm matter}}{\delta \Phi^2}=0\,. 
\label{class-s}
\end{eqnarray}
This case is very special because the equations of motion for $\omega_1$ 
and $\omega_2$ remain to be a pair of Liouville equations because 
the right-hand sides of the first and second equations in (\ref{matter eq}) vanish. 
Hence one can still use the general solutions (\ref{Liouville}). 
The constraints are also still written in terms of Schwarzian derivatives, 
but slightly modified like  
\begin{eqnarray}
{\rm Sch}\{ X^\pm_1, x^\pm\} - {\rm Sch}\{ X^\pm_2, x^\pm\} 
&=& -32 \pi G\, \eta\, T_{\pm\pm}\,. 
\end{eqnarray}
That is, the right-hand side does not vanish. 

\medskip 

To solve the set of equations, it is helpful to introduce new functions 
$\varphi_\pm=\varphi_\pm(x^\pm)$ defined as 
\begin{eqnarray}
\varphi_\pm \equiv \frac{1}{ \sqrt{| \partial_\pm X^\pm_2 |} }\,. 
\end{eqnarray}
Note here that $X_2^{\pm}$ only have been utilized. 
Then by using $\varphi_\pm$\,, the Schwarzian derivatives can be rewritten as 
\begin{eqnarray}
{\rm Sch}\{ X^\pm_2, x^\pm \} 
= -2 \frac{ \partial_\pm^2 \varphi_\pm }{ \varphi_\pm }\,. 
\end{eqnarray}
When the coordinates are taken as 
\begin{eqnarray}
X^\pm_1(x^\pm) = x^\pm\,, 
\end{eqnarray}
the constraints become Schr$\ddot{\rm o}$dinger equations as follows: 
\begin{eqnarray}
\left(-\partial_\pm^2 -16 \pi G \eta T_{\pm\pm} \right)\varphi_{\pm}(x^\pm) &=&0\,. 
\end{eqnarray}
Thus, for the simple class of matter fields, the constraints have been drastically simplified. 

\subsection{A solution describing formation of a black hole}

As an example in the simple class, let us consider an ingoing matter pulse of energy 
$E/(8\pi G)$: 
\begin{eqnarray}
T_{--} &=& \frac{E}{8 \pi G}\, \delta(x^-)\,, \qquad T_{++}=T_{+-} = 0\,. 
\end{eqnarray}
Note here that $T_{\mu\nu}$ does not depend on the dilaton $\Phi^2$ and hence 
this case belongs to the simple class (\ref{class-s})\,. 
This pulse causes a shock-wave traveling on the null curve $x^-=0$\,. 

\medskip

Then the constraint for the anti-holomorphic part is written as 
\begin{eqnarray}
\left(-\partial_-^2 -2E\eta  \delta(x^-) \right)\varphi_-(x^-) &=&0\,. 
\end{eqnarray}
By solving this equation, we obtain the following solution: 
\begin{eqnarray}\label{sol-phi}
\varphi_-(x^-) &=& \varphi_-(0) \left(1-2E \eta x^- \theta(x^-) \right)\,. 
\end{eqnarray}
Assuming the continuity, $X^-_2$ is given by\footnote{
Note here that for simplicity, we dropped and tuned some integration constants in \eqref{sol-phi} and \eqref{sol-xm}.}
\begin{eqnarray}\label{sol-xm}
X^-_2(x^-)
&=&\left\{\begin{array}{ll}
(1-4 \eta^2 E a)x^- -2 \eta a & \quad ({\text{for}}~~x^-<0) \\
\frac{x^--2 \eta a}{ 1 -2 E \eta x^- } & \quad ({\text{for}}~~x^->0) \\
\end{array} \right.\,.
\end{eqnarray}
Here $a$ is an arbitrary integral constant and the scaling factor $\varphi_-(0)$ is fixed as
\begin{eqnarray}
\varphi_-(0)^2 =\frac{1}{1-4\eta^2 E a}\,. 
\end{eqnarray}

\medskip 

The remaining task is to determine $X_2^+(x^+)$\,. The constraint for $\varphi_+(x^+)$ 
is given by
\begin{eqnarray}
-\partial_+^2 \varphi_+(x^+) =0\,.
\end{eqnarray} 
Thus one can determine $\varphi_+(x^+)$ and $\partial_+ X_2^+(x^+)$ as 
\[
\varphi_+(x^+) = \gamma x^+ + \delta\,, \qquad 
\partial_+ X_2^+(x^+) = \frac{1}{(\gamma x^+ + \delta)^2}\,, 
\]
where $\gamma$ and $\delta$ are constants. Hence $X_2^{+}$ is obtained as 
\begin{eqnarray}
X_2^+ (x^+) = \frac{\alpha x^+ + \beta}{\gamma x^+ + \delta} \qquad 
(\alpha \delta -\beta \gamma = 1)
\end{eqnarray}
with new constants $\alpha$ and $\beta$\,. For simplicity, we will set 
$\alpha=\delta=1,~\beta=\gamma=0$\,. That is, $X_2^+ = x^+$\,. 

\medskip

Thus one can obtain a solution of the two Liouville equations as follows: 
\begin{eqnarray}
{\rm e}^{2 \omega_1 } &=&\frac{4}{ \left( x^+  - x^-\right)^2}\,, \no\\
{\rm e}^{2 \omega_2 }
&=&
\left\{\begin{array}{ll}
\frac{4(1-4 \eta^2 E a) }{  \left( x^+ -x^-+2\eta a(1 + 2 \eta E  x^-)  \right)^2} 
& \quad ({\text{for}}~~x^-<0) \\
\frac{4(1-4 \eta^2 E a) }{  \left( x^+ - x^-+2\eta(a - E  x^+ x^-) \right)^2} 
& \quad ({\text{for}}~~x^->0) 
\end{array} \right.\,
\,. 
\end{eqnarray}
As a result, the original metric and dilaton are given by 
\begin{eqnarray}
{\rm e}^{2\omega}  
&=&
\left\{\begin{array}{ll}
\frac{4\sqrt{1-4 \eta^2 E a} }{  \left(x^+-x^- \right) \left(x^+ -x^-+2\eta a(1 +2 \eta E  x^- \right) } 
& \quad ({\text{for}}~~x^-<0) \\
\frac{4\sqrt{1-4 \eta^2 E a} }{  \left(x^+-x^- \right) \left( x^+ - x^-+2\eta(a - E  x^+ x^-) \right)} 
& \quad ({\text{for}}~~x^->0) 
\end{array} \right.\,\,, \no\\
\Phi^2  
&=&
\left\{\begin{array}{ll}
\frac{1}{2 \eta} \log \left| 1+ \frac{ 2\eta a(1+ 2 E \eta x^-)}{  x^+ - x^- } \right|\, 
-\frac{1}{2\eta}\log(1-4\eta^2Ea)& \quad ({\text{for}}~~x^-<0) \\
\frac{1}{2 \eta} \log \left| 1+ \frac{ 2\eta(a- E x^+x^-)}{  x^+ - x^- } \right|\, 
-\frac{1}{2\eta}\log(1-4\eta^2Ea) & \quad ({\text{for}}~~x^->0) 
\end{array} \right.\,. 
\end{eqnarray}
The undeformed limit $\eta\to 0$ leads to a solution describing formation 
of a black hole in the undeformed model \cite{AP}. 
Note here that the energy-dependent constant in $\Phi^2$ vanishes 
in the undeformed limit. At least so far, we have no idea 
for the physical interpretation of this constant.

\section{The deformed system with a conformal matter}

In this section, we will consider conformal matters, which do not belong 
to the previous class (\ref{class-s}), and discuss the effect of them to 
thermodynamic quantities associated with a black hole solution.  

\medskip 

Let us study a conformal matter whose dynamics is governed 
by the classical action: 
\begin{eqnarray}
S_{\rm matter} &=&  -\frac{N}{ 24 \pi }\int\!\dd^2 x\, 
\sqrt{-g}\,\Bigl[\chi ( R -2 \eta \nabla^2  \Phi^2) 
+ (\nabla \chi)^2  \Bigr]
-\frac{N}{12 \pi }\int \dd t\, \sqrt{-\gamma_{tt} }\,K\,.
\end{eqnarray}
Here $N$ denotes the central charge of $\chi$\,. 
It is worth noting that the conformal matter couples to dilaton as well as the Ricci scalar, 
in comparison to the undeformed case \cite{AP}.  
Then the energy-momentum tensor and a variation of $S_{\rm matter}$ 
with respect to the dilaton are given by 
\begin{eqnarray}
T_{+-}&=&\frac{N}{12 \pi}\partial_+\partial_-\chi\,, \no\\
T_{\pm\pm}&=&\frac{N}{12 \pi}(-\partial_\pm \partial_\pm \chi 
+ \partial_\pm\chi \partial_\pm\chi +2\partial_\pm \chi \partial_\pm \omega_1)\,, \no\\
\frac{\delta S_{\rm matter}}{\delta \Phi^2}&=&-\frac{N}{6 \pi}\eta \,\partial_+\partial_-\chi\,. 
\end{eqnarray}
Hence the equations of motion are given by 
\begin{eqnarray}
 \partial_+ \partial_-(\omega_1 + \chi) &=& 0\,, \no\\
4\partial_+\partial_- \omega_{1} + {\rm e}^{2\omega_{1} } 
&=&\frac{16}{3}GN  \eta \partial_+ \partial_- \chi \,, \no\\
4\partial_+\partial_- \omega_{2} + {\rm e}^{2\omega_{2} } 
&=&0 \,, \no\\
{\rm e}^{\omega_{1} } \partial_{\pm}\partial_{\pm} {\rm e}^{ -\omega_{1} } 
- {\rm e}^{\omega_{2} } \partial_{\pm}\partial_{\pm} {\rm e}^{ -\omega_{2} } 
&=& \frac{2}{3}GN (-\partial_{\pm} \partial_{\pm} \chi + \partial_{\pm} \chi  \partial_{\pm} \chi  
+2  \partial_{\pm} \chi  \partial_{\pm} \omega_{1} )\,. \label{confo}
\end{eqnarray}
Note here that the third equation is still the Liouville equation, 
while the second equation acquired the source term due to the matter contribution. 

\medskip 

As we will see below, the system of equations (\ref{confo}) is still tractable and 
one can readily find out a black hole solution including the back-reaction 
from the conformal matter $\chi$\,.

\subsection{A black hole solution with a conformal matter}

Let us derive a black hole solution. 

\medskip

Given that the solution is static, $\chi$ can be expressed as 
\begin{eqnarray}
\chi = -\omega_1-\sqrt{\mu}\,(x^+-x^-) \,.
\end{eqnarray}
By eliminating $\chi$ from the other equations, one can derive a couple of Liouville equations 
and the constraint conditions:
\begin{eqnarray}
4\left(1+\frac{4}{3}GN \eta \right)\partial_+\partial_- \omega_{1} 
+ {\rm e}^{2\omega_{1} } &=&0\,, \no\\
4\partial_+\partial_- \omega_{2} + {\rm e}^{2\omega_{2} } &=&0 \,, \no\\
\left(1+\frac{2}{3}GN \right) {\rm e}^{\omega_{1} } \partial_{\pm}\partial_{\pm} 
{\rm e}^{ -\omega_{1} } - {\rm e}^{\omega_{2} } \partial_{\pm}\partial_{\pm} 
{\rm e}^{ -\omega_{2} } &=&\frac{2}{3}GN\mu \,. 
\end{eqnarray}
Note that a numerical coefficient in the first equation is shifted by a certain constant 
as a non-trivial contribution of the conformal matter. 

\medskip 
 
Still, we can use the general solutions of Liouville equations given by 
\begin{eqnarray}
{\rm e}^{2 \omega_{1} }&=&  \frac{ 4  ( 1+\frac{4}{3}GN \eta ) }{  
\left( X^+_1 - X^-_1 \right)^2} \partial_+ X^+_1 \partial_- X^-_1\,,  \no\\
{\rm e}^{2 \omega_{2} }&=&   \frac{ 4  }{  \left( X^+_2 - X^-_2 \right)^2}
\partial_+ X^+_2 \partial_- X^-_2\,. 
\end{eqnarray}
By using $X^\pm_i~(i=1,2)$ and the Schwarzian derivative, the constraints can be rewritten as 
\begin{eqnarray}
\left(1+\frac{2}{3}GN \right){\rm Sch}\{ X^+_1, x^+\} 
- {\rm Sch}\{ X^+_2, x^+\}  &=& -\frac{4}{3}GN\mu\,, \no \\ 
\left(1+\frac{2}{3}GN\right) {\rm Sch}\{ X^-_1, x^-\} - {\rm Sch}\{ X^-_2, x^-\}  
&=& -\frac{4}{3}GN\mu\,. \label{cons}
\end{eqnarray}
It is an easy task to see that the hyperbolic-type coordinates 
\begin{eqnarray}
X^\pm_{1,2} &=&L^\pm_{1,2}\left( \tanh(\sqrt{\mu} x^\pm)\right)\,, \label{coor}
\end{eqnarray}
satisfy the constraints (\ref{cons})\,, where $L_{1,2}^\pm$ denote 
linear fractional transformations as in \eqref{LFT}.
Note that each of $X_{1,2}^\pm$ covers a partial region of the original spacetime. 
Hence the coordinate transformations (\ref{coor}) may lead to a black hole solution \cite{AP, KOY}. 
In fact, the Schwarzian derivatives have particular values like  
\begin{eqnarray}
{\rm Sch}\{ L_{1,2}^\pm \left(\tanh(\sqrt{\mu}\, x^\pm )\right),\, x^\pm \} = -2\mu \,, 
\end{eqnarray}
and hence these coordinates satisfy the constraints.

\medskip 

Here we choose the following linear transformations $L_{1,2}^\pm$:
\begin{eqnarray}
X^+_1(x^+) &=& 
\frac{ \tanh(\sqrt{\mu}\,x^+) - \eta \sqrt{\mu} }{- \eta \mu \tanh(\sqrt{\mu}\,x^+) + \sqrt{\mu} } 
\,, \no\\ \no\\
X^+_2(x^+) &=& 
\frac{ \tanh(\sqrt{\mu}\,x^+) + \eta \sqrt{\mu}  }{\eta \mu \tanh(\sqrt{\mu}\,x^+) +\sqrt{\mu} }  
\,,  \no\\ \no \\
X^-_1(x^-) &=&   X^-_2(x^-) =  \frac{1}{ \sqrt{\mu} } \,\tanh(\sqrt{\mu} x^-)\,, 
\end{eqnarray}
one can derive a deformed black hole solution with conformal matters: 
\begin{eqnarray}
{\rm e}^{2 \omega }&=&\frac{ 4\mu (1-\eta^2\mu)\sqrt{1+\frac{ 4}{3}GN \eta} }{  
\sinh^2(2 \sqrt{\mu}\,Z )-\eta^2\mu \cosh^2(2\sqrt{\mu}\,Z)}\,, 
\label{dBHmetric} \\
\Phi^2 &=& \frac{1}{2 \eta} \log \left|
\frac{1+\eta\sqrt{\mu}\, \coth(2\sqrt{\mu}\,Z)}{1-\eta\sqrt{\mu}\, 
\coth(2\sqrt{\mu}\, Z)}\right|+\frac{1}{4\eta}\log\left(1+\frac{ 4}{3}GN \eta \right)\,.  
\label{dBHdilaton}
\end{eqnarray}

The matter effect just changes the overall factor of the metric and shifts the dilaton 
by a constant. In the undeformed limit $\eta \to 0$\,, this solution reduces to 
a black hole solution with conformal matters presented in \cite{AP}: 
\begin{eqnarray}
{\rm e}^{2 \omega } =\frac{ 4\mu }{  \sinh^2(2 \sqrt{\mu}\,Z )}\,, \qquad 
\Phi^2 =
\sqrt{\mu}\, \coth(2\sqrt{\mu}\,Z)+\frac{1}{3}GN\,. 
\end{eqnarray}

\subsection{Black hole entropy}

In this subsection, we shall compute the entropy of the black hole solution with a conformal matter 
given in (\ref{dBHmetric}) and (\ref{dBHdilaton}) from two points of view: 
1) the Bekenstein-Hawking entropy and 2) the boundary stress tensor with a certain counter-term. 

\subsubsection*{1) the Bekenstein-Hawking entropy }

Let us first compute the Bekenstein-Hawking entropy. 
From the metric  (\ref{dBHmetric}), one can compute the Hawking temperature as 
\begin{eqnarray}
T_{\rm H}=\frac{ \sqrt{\mu} }{\pi}\,.  
\label{HT}
\end{eqnarray}
From the classical action, the effective Newton constant $G_{\rm eff}$ is determined as  
\begin{eqnarray}
\frac{1}{G_{\rm eff}} = \frac{\Phi^2}{G} - \frac{2}{3}N \chi \,. 
\label{Geff}
\end{eqnarray}
Note that the presence of the conformal matter fields is reflected as a shift of $G_{\rm eff}$\,. 
Given that the horizon area A is $1$\,,  
the Bekenstein-Hawking entropy $S_{\rm BH}$ is computed as 
\begin{eqnarray}
S_{\rm BH} 
&=& \left. \frac{A}{4 G_{\rm eff} }\right|_{Z \to \infty} \,\no\\
&=& \frac{ 1+\frac{2}{3}GN\eta }{4 G \eta} \text{arctanh}{ ( \pi \,T_{\rm H} \, \eta)} 
+\frac{N}{6}\log(T_{\rm H}) \no\\
&& + \frac{1+\frac{4}{3}G N \eta }{ 16 G \eta} \log\left( 1+\frac{4}{3}GN\eta \right) +\frac{N}{6}\log(4\pi)
\,.\label{BH}
\end{eqnarray}
The terms in the last line are constants independent of the Hawking temperature.

\subsubsection*{2) the boundary stress tensor }

The next is to evaluate the entropy by computing the boundary stress tensor 
with a certain counter-term. 

\medskip

In conformal gauge, the total action including the Gibbons-Hawking term
can be rewritten as
\begin{eqnarray}
S_{\Phi} 
&=&\frac{1}{8 \pi G}\int\!\dd^2 x \left[
-4 \partial_{(+} \Phi^2 \partial_{-)} \omega  
+ \frac{1}{2\eta}\sinh(2\eta\Phi^2)\, {\rm e}^{2 \omega} \right]\,, \no\\
S_{\rm matter}
&=&\frac{N}{6 \pi}\int\dd^2 x \,
\Bigl[ \partial_+\chi\partial_-\chi+2\partial_{(+}\chi\partial_{-)}
\omega+2\eta\partial_{(+}\chi\partial_{-)}\Phi^2
\Bigr]\,.  
\end{eqnarray} 
By using the explicit expression of the black hole solution in (\ref{dBHmetric}) and (\ref{dBHdilaton}), 
the on-shell bulk action can be evaluated on the boundary, 
\begin{eqnarray}
S_{\Phi} + S_{\rm matter} &=&
 \int \! \dd t\,  \left. \frac{ -(1+\frac{2}{3}GN \eta)\mu 
 -\frac{\sqrt{\mu}GN }{3}(1-\eta^2\mu)\sinh(4\sqrt{\mu}Z) }
{2\pi G ( 1+\eta^2 \mu +(-1+\eta^2 \mu )
 \cosh(4\sqrt{\mu} Z) ) } \right|_{Z\rightarrow Z_0}\,. \label{bulk}
\end{eqnarray}
As argued in \cite{KOY}\footnote{For an earlier argument for the relation between the singularity and 
the holographic screen, see \cite{Kame}.}, 
the singularity of (\ref{dBHmetric}) is identified as the boundary $Z_0$: 
\begin{eqnarray}
Z_0 \equiv \frac{1}{2\sqrt{\mu}} \rm{arctanh}(\eta\sqrt{\mu})\,.
\end{eqnarray}
As the bulk action approaches the boundary $(Z \to Z_0)$\,, the bulk action (\ref{bulk}) diverges 
and hence one needs to introduce a cut-off. 
When the regulator $\epsilon$ is introduced such that $Z-Z_{0}=\epsilon$\,, 
the on-shell action is expanded as 
\begin{eqnarray}
S_{\Phi} + S_{\rm matter} &=& \int \! \dd t\, \left[
\frac{1+\frac{4}{3}GN\eta }{ 16 \pi G\, \eta\, \epsilon  } - \frac{1+\eta^2 \mu }{16 \pi G\, \eta^2}\,
 +O(\epsilon^1) \right]\,. 
\end{eqnarray}

\medskip 

To cancel the divergence, 
it is appropriate to add the following counter-term:\footnote{
Note that in the undeformed limit $\eta\rightarrow 0$, this counter-term reduces to 
the one in \cite{AP}. When $\mu=N=0$, this term becomes the dilaton potential 
$\frac{1}{\eta}\sinh (2 \eta \Phi^2)$.}
\begin{eqnarray}
S_{\rm ct} &=& \frac{-1}{8 \pi G}\int\!\dd t\, \frac{\sqrt{-\gamma_{tt}} }{L}\, \left[
\sqrt{ F \left( \Phi^2 \right) - \frac{1}{ \eta^2 } \log(1-\eta^2 \mu)} \right. \no \\ 
&& \hspace*{3.5cm} \left. +\frac{4}{3}GN\sqrt{G(\Phi^2)-\frac{1}{2\eta^2}\log(1-\eta^2\mu)}\,
\right]\,. 
\label{counter}
\end{eqnarray}
Here $L$ is the overall factor of the metric defined as 
\begin{eqnarray}
L^2 \equiv  (1-\eta^2\mu)\sqrt{1+\frac{ 4}{3}GN \eta}\,, 
\end{eqnarray}
and scalar functions $F$ and $G$ are defined as 
\begin{eqnarray}
 F \left( \Phi^2 \right) &\equiv& 
\frac{ -1-\eta^2 \mu +(1-\eta^2 \mu) \cosh\left[ 2\eta\left( \Phi^2 
- \frac{ \log\left( 1+\frac{4}{3}GN\eta \right)}{4\eta} \right) \right]}{ 2 \eta^2 }\,,  \no\\
G(\Phi^2)&\equiv&
\frac{ 8 \eta \sqrt{\mu} +(1-\eta^2 \mu) 
\exp\left[ 2\eta\left( \Phi^2-\frac{ \log\left( 1+\frac{4}{3}GN\eta \right)}{4\eta} \right) 
\right]}{ 4 \eta^2 }\,. 
\end{eqnarray}
The extrinsic metric $\gamma_{tt}$ on the boundary is evaluated as 
\[
\gamma_{tt} =\left. - {\rm e}^{2 \omega} \right|_{Z \to Z_{0}} \,. 
\]
In the undeformed limit $\eta \to 0$, this counter-term reduces to
\begin{eqnarray}
S_{\rm ct}^{(\eta =0)} &=& \int\!\dd t\, \sqrt{-\gamma_{tt}}\, 
\left( -\frac{\Phi^2}{8 \pi G} -\frac{N}{24 \pi }\right)\,. 
\end{eqnarray}
This is nothing but the counter-term utilized in the undeformed model \cite{AP}.

\medskip

It is straightforward to check that the sum $S = S_{\Phi}+S_{\rm matter}+S_{\rm ct}$ 
becomes finite on the boundary by using the expanded form of the counter-term (\ref{counter}): 
\begin{eqnarray}
S_{\rm ct} &=& \int \! \dd t\, \left[
 -\frac{1+\frac{4}{3}GN\eta }{16 \pi G \eta\, \epsilon } + \frac{1+\eta^2 \mu 
 -\frac{16}{3}GN\eta^2+2( 1+\frac{2}{3}GN\eta ) 
 \log(1-\eta^2 \mu) }{ 16 \pi G\, \eta^2}\, +O(\epsilon) \right]\,.\no
\end{eqnarray}

\medskip

In a region near the boundary, the warped factor of the metric (\ref{dBHmetric}) is expanded as 
\begin{eqnarray}
 {\rm e}^{2 \omega} &=& \frac{L}{ \eta\, \epsilon } + O(\epsilon^0)\,. 
\end{eqnarray}
Hence, by normalizing the boundary metric as 
\[
\hat{\gamma}_{tt}  =\frac{ \eta\, \epsilon}{L} \, \gamma_{tt}\,, 
\]
the boundary stress tensor is defined as 
\begin{eqnarray}
\langle \hat{T}_{tt} \rangle 
&\equiv& \frac{-2}{\sqrt{-\hat{\gamma}_{tt}}}\, \frac{\delta S}{\delta \hat{\gamma}^{tt} } 
= \lim_{ \epsilon \to 0 } \sqrt{ \frac{\eta\, \epsilon}{L} }\, \frac{-2}{\sqrt{-\gamma_{tt} } }\, 
\frac{\delta S}{\delta \gamma^{tt} }\,.  
\end{eqnarray}
After all, $\langle \hat{T}_{tt} \rangle$ is evaluated as  
\begin{eqnarray}
\langle \hat{T}_{tt} \rangle &=& 
\frac{ -(1+\frac{2}{3}GN\eta )\log(1-\eta^2 \mu) }{8 \pi G\, \eta^2}
+\frac{N \sqrt{\mu}}{6\pi}\,. 
\end{eqnarray}

\medskip

To compute the associated entropy, 
$\langle \hat{T}_{tt} \rangle$ should be identified with energy $E$ like 
\begin{eqnarray}
E &=& \frac{ -(1+\frac{2}{3}GN\eta)\log(1-\pi^2 T_{\rm H}^2 \eta^2 ) }{8 \pi G\, \eta^2}
+\frac{N}{6}T_{\rm H}\,,
\end{eqnarray}
where we have used the expression of the Hawking temperature (\ref{HT})\,. 
Then by solving the thermodynamic relation,
\begin{eqnarray}
\dd E &=&\frac{\dd S}{T_{\rm H} }\,,
\end{eqnarray}
the associated entropy is obtained as  
\begin{eqnarray}
S &=& \frac{ (1+\frac{2}{3}GN\eta) }{ 4 G \eta }\text{arctanh}{(\pi T_{\rm H} \eta)} 
+ \frac{N}{6}\log(T_{\rm H})+S_{T_{\rm H}=0}\,. 
\end{eqnarray}
Here $S_{T_{\rm H}=0}$ has appeared as an integration constant that measures 
the entropy at zero temperature. Thus the resulting entropy precisely 
agrees with the Bekenstein-Hawking entropy (\ref{BH})\,, 
up to the temperature-independent constant.

\section{Conclusion and discussion}

In this paper, we have considered some matter contributions 
to a (1+1)-dimensional dilaton gravity system with a hyperbolic dilaton potential. 
By introducing a couple of new variables, this system has been rewritten into  
a pair of Liouville equations with two constraints. In particular,  
the constraints in conformal gauge can be expressed in terms of Schwarzian derivatives. 
We have revisited the vacuum solutions and revealed its dipole-like structure. 
The new variables are so powerful in studying solutions. As a benefit, 
we have constructed a time-dependent solution which describes formation of a black hole 
with a pulse. Finally, the black hole entropy has been considered 
by taking account of conformal matters. 
The Bekenstein-Hawking entropy agrees with the entropy computed from the boundary 
stress tensor with a certain counter-term.  

\medskip 

There are some future directions. The first is to clarify a connection   
between the system considered here and the doubled formalism such as  
Double Field Theory (DFT) \cite{Siegel,HZ,HHZ} 
and Double Sigma Model (DSM) \cite{Tseytlin,Hull,Copland,Park}. 
As well recognized, Yang-Baxter deformations of type IIB string 
theory defined on AdS$_5\times$S$^5$ \cite{DMV2,KMY-Jordanian-typeIIB} 
are closely related to DFT and DSM \cite{SUY,BMS,SSY} 
via the generalized supergravity \cite{AFHRT,WT}. 
A similar connection may be expected in the present lower-dimensional case as well, because 
the present system was originally constructed by employing the Yang-Baxter deformation technique. 
The second is to reveal the underlying symmetry.  
By following a nice work by Ikeda and Izawa \cite{II}, the hyperbolic dilaton potential 
leads to the expected $q$-deformed $\mathfrak{sl}(2)$ algebra realized in the associated non-linear 
gauge theory. Elaborating this symmetry algebra helps us to identify the holographic dual. 
The third is to a generalization to include arbitrary matter fields 
and discuss the associated one-dimensional boundary theory by following \cite{MSY,EMV}. 
It seems likely that the anticipated system is a deformed Schwarzian theory. 
Finally, it is interesting to consider a similar deformation of the asymptotically flat case 
\cite{CGHS} by following \cite{MORSY}. It is also nice to study 
how the holographic relation should be modified in the case with a reflecting dynamical boundary 
by generalizing \cite{Yegor}. The integrability techniques discussed there would still be  
useful even after performing Yang-Baxter deformations. 

\medskip 

We hope that the dipole-like structure uncovered here would shed light 
on a new aspect of the 2D dilaton gravity system and further the holographic principle as well.

\subsection*{Acknowledgments}

We are very grateful to Noriaki Ikeda and Benoit Vicedo for useful comments and discussions.  
The work of H.K. is supported by the Japan Society for the Promotion of Science (JSPS).
The work of K.Y. is supported by the Supporting Program for Interaction-based Initiative Team Studies 
(SPIRITS) from Kyoto University and by a JSPS Grant-in-Aid for Scientific Research (C) No.\,15K05051.
This work is also supported in part by the JSPS Japan-Russia Research Cooperative Program.

\end{document}